\documentclass[sn-nature]{sn-jnl}

\usepackage{graphicx}%
\usepackage{multirow}%
\usepackage{amsmath,amssymb,amsfonts}%
\usepackage{amsthm}%
\usepackage{mathrsfs}%
\usepackage[title]{appendix}%
\usepackage{xcolor}%
\usepackage{textcomp}%
\usepackage{manyfoot}%
\usepackage{booktabs}%
\usepackage{algorithm}%
\usepackage{algorithmicx}%
\usepackage{algpseudocode}%
\usepackage{listings}%
\usepackage{mathtools}
\usepackage{siunitx}
\usepackage{lineno}

\DeclarePairedDelimiterXPP\BigOSI[2]%
  {\mathcal{O}}{(}{)}{}%
  {\SI{#1}{#2}}

\begin{document}

\title[Article Title]{
Origins of complexity in the rheology of Soft Earth suspensions
}

\author[1,2]{\fnm{Shravan} \sur{Pradeep}}\email{spradeep@sas.upenn.edu}

\author[2]{\fnm{Paulo E.} \sur{Arratia}}\email{parratia@seas.upenn.edu}

\author*[1,2]{\fnm{Douglas J.} \sur{Jerolmack}}\email{sediment@sas.upenn.edu}

\affil[1]{\orgdiv{Department of Earth and Environmental Science}, \orgname{University of Pennsylvania}, \orgaddress{\city{Philadelphia}, \postcode{19104}, \state{PA}, \country{USA}}}

\affil[2]{\orgdiv{Department of Mechanical Engineering and Applied Mechanics}, \orgname{University of Pennsylvania}, \orgaddress{\city{Philadelphia}, \postcode{19104}, \state{PA}, \country{USA}}}

\abstract{When wet soil becomes fully saturated by intense rainfall, or is shaken by an earthquake, it may fluidize catastrophically. Sand-rich slurries are treated as granular suspensions, where the failure is related to an unjamming transition. Mud flows are modeled as gels, where yielding and shear-thinning behaviors arise from inter-particle attraction and clustering. Here we show that the full range of complex flow behaviors previously reported for natural debris flows can be reproduced with three ingredients: water, silica sand, and kaolin clay. Going from sand-rich to clay-rich suspensions, we observe continuous transition from brittle to ductile yielding. We propose a general constitutive relation for soil suspensions, with a particle rearrangement time that is controlled by yield stress and jamming distance. Our experimental results are supported by models for amorphous solids, suggesting that the paradigm of non-equilibrium phase transitions can help us understand and predict the complex behaviors of Soft Earth suspensions.}

\keywords{dense suspensions, Soft Earth, geophysical flows, brittle-to-ductile failure}

\maketitle

\section*{Introduction}

Earth's surface is covered in heterogeneous and soft particulate material that we call soil. To first order, we may consider Soft Earth suspensions -- and their associated geophysical flows -- to be composed of three ingredients: frictional particles (sand and/or silt), cohesive particles (clay, fine silt, organic materials etc.), and water \cite{jerolmack2019viewing}. Apparently solid soil can suddenly lose rigidity when stressed; for example, earthquake-induced liquefaction, the collapse of earth dams, or the failure of water-soaked hillsides that forms landslides and debris flows (Fig. 1a) \cite{iverson1997debris, huang2017hazard}. Whether and how soil yields and flows under environmental loads is exquisitely sensitive to the relative proportions of the three main ingredients \cite{huang2018effects}. Here we focus on debris flows: dense, viscous and fast-flowing suspensions of soil. With the increasing frequency and intensity of extreme weather and wildfire due to climate change, the hazard posed by debris flows is growing \cite{kean2021forecasting}. Debris flows are complex and transient phenomena, where shear stress, flow speed, particle size and water content co-evolve \cite{iverson2014depth}. Controlled laboratory tests are useful for taming this complexity by fixing variables, allowing determination of the constitutive relations among shear stress ($\tau$), solids volume fraction ($\phi$) and shear rate ($\dot{\gamma}$). Measured `flow curves' are typically fit with the phenomenological Hershel-Bulkley equation, $\tau = \tau_y + k \dot{\gamma}^n$, where $\tau_y$ is a yield stress below which flow cannot occur, $n$ is the flow index and $k$ is an empirical constant. Previous studies have established that debris flows are typically shear thinning ($n<1)$ and have a significant yield stress ($\tau_y >0$), and that these properties are sensitive to the particle concentration \cite{major1992debris, coussot1994behavior, schippa2020modeling}. Even under idealized conditions, however, laboratory studies have not provided a clear picture on how debris-flow material composition controls flow behavior. The physical meaning of $k$ and $n$ are ambiguous for natural debris slurries, which limits extrapolation of results beyond the laboratory.

Some physical insight has been gained by classifying debris flows into two end-member types based on soil composition: ``Granular'' and ``Cohesive'' \cite{coussot1996recognition, ancey2007plasticity}. Granular debris flows are composed mostly of frictional, cohesionless soil -- i.e, sand and larger particles. Our previous work demonstrated that weakly cohesive debris-flow materials from the California Coast Range could be described by constitutive relations developed for idealized granular suspensions \cite{boyer2011unifying}, by accounting for two material-specific quantities associated with jamming of granular media: the jamming fraction ($\phi_m$) above which flow cannot occur, and the suspension yield stress \cite{kostynick2022rheology}.  Using these quantities to normalize the data, we found the resultant flow behavior of debris-flow materials was well described by a dimensionless Bingham model: elastic-like behavior at low shear rate/stress, viscous flow behavior (i.e., $n=1$) at high shear rate/stress, and a rapid transition between them consistent with discontinuous, Coulomb-like failure (Fig. 1b). The apparent variations in $n$, observed in the raw flow curves, were eliminated once the shear stresses and shear rates for each suspension were properly normalized; we refer the reader to the paper for details \cite{kostynick2022rheology}. This picture is incompatible, however, with the Cohesive debris-flow end member. Clay-rich debris-flow materials from the French Alps were shown to yield gradually and continuously, with $n$ much smaller than one \cite{coussot1995structural} -- flow curves cannot be fit with the same Bingham model (Fig. 1b). Yielding dynamics of clay suspensions are strongly related to attractive interactions among particles, and these suspensions may exhibit solid-like behavior even at low $\phi$ \cite{ran2023understanding,coussot2002avalanche,bonn2002rheology}. Clay suspensions also exhibit hysteresis and thixotropy, that change with the strength of the inter-particle attractions (through particle surface charge and solution ionic strength) \cite{bonn2017yield, ran2023understanding}. All of these behaviors are reminiscent of gels, in which inter-particle attraction and clustering -- rather than friction and jamming -- determine their yield and flow properties \cite{bonn2017yield}. 

Motivated by industrial applications, there is increasing interest in examining the rheologic behaviors of granular materials suspended in complex fluids such as gels \cite{jiang2022flow,li2023impact,jiang2023colloidal, kammer2022homogenization} and emulsions \cite{dagois2015rheology}. While progress has been made in extending constitutive models to such complex mixtures \cite{dagois2015rheology, kammer2022homogenization}, no study has probed the transition from granular suspension to gel. Some insight on the possible nature of this transition can be gained by examining yielding in other amorphous solids. A numerical study discovered that a transition from brittle yielding typical of ``hard'' glass materials, to ductile yielding common in ``soft'' glassy materials, can arise within an amorphous solid by changing its degree of annealing -- i.e., the initial preparation and stability of the material \cite{ozawa2018random}. Theoretical analysis of elastoplastic models -- which are commonly used to describe soft glassy materials like emulsions, gels and foams -- has constrained the value $n \approx 1/2$ in the vicinity of yielding \cite{bocquet2009kinetic, lin2018microscopic}. The reported rheologies of debris-flow materials summarized earlier indicate that there must be a transition from frictional to cohesive control, as the relative proportion of sand to clay is increased in a suspension \cite{ancey2001role,ancey2001yield,ancey2007plasticity}. Flow curves of the two end members suggest that this corresponds to a transition from brittle, Coulomb-type failure to ductile, plastic failure (Fig. 1b). Here we build a minimal experimental system to examine this transition: suspensions of deionized water, frictional silica sand, and cohesive kaolin clay. Carefully controlled rheologic tests reveal a continuous transition from brittle to ductile yielding as the relative proportion of clay to sand is increased. The transition appears to be related to percolation of cohesive-particle networks. Our model Soft Earth suspensions capture the full range of previously reported debris-flow rheologies (Fig. 1b), and are used to build a physically-based constitutive relation with parameters that are related to material properties.   

\begin{figure}[h]
    \centering
    \includegraphics[width=1.05\linewidth]{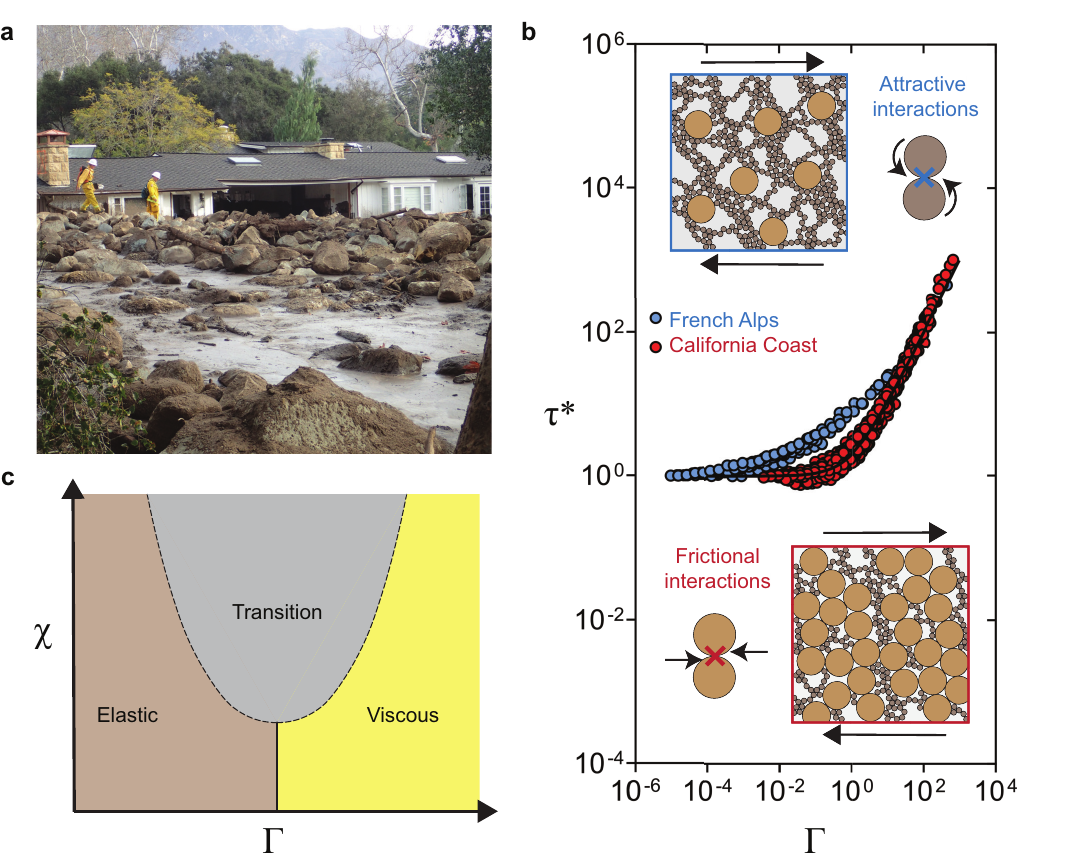}
    \caption{\textbf{Debris flows and rheological phase space. (a)} Debris flow deposit at Montecito, in the California Coast Range (2018; Photo Credit: United States Geological Survey). \textbf{(b)} Non-dimensionalized flow curves of debris-flow materials from the French Alps (blue circles \cite{coussot1995structural}) and the California Coast Range (red circles \cite{kostynick2022rheology}) on a scaled shear stress ($\tau^\star$) - shear rate ($\Gamma$) plot. A simplified Bingham model, $\tau^\star = 1 + \Gamma$, is fit to the California Coast Range data but cannot fit the French Alps data. We hypothesize that stress transmission in the former is dominated by granular friction, while in the latter it is governed by cohesion. Cohesive particles are expected to impose a rotational constraint on motion, while frictional particles impart a sliding constraint \cite{guy2018constraint}. \textbf{(c)} Proposed phase space for Soft Earth suspensions defined using scaled variables: interparticle attraction strength ($\chi$) on the y-axis, and non-dimensional shear rate ($\Gamma$) on the x-axis. Low $\chi$ suspensions are expected to exhibit brittle yielding; as $\chi$ increases, a transitional plastic regime (rate-dependent plasticity) emerges and grows as yielding becomes increasingly ductile. Error bars indicate SD.}     
\end{figure}

\section*{Results}

\subsection*{Developing a generic constitutive relation}
 
The dimensionless Bingham model that we previously developed for weakly cohesive debris-flow slurries \cite{kostynick2022rheology} has the form $\tau^\star = 1 + \Gamma$, where the dimensionless shear stress is $\tau^\star \equiv \tau/\tau_y$, and the dimensionless viscous stress $\Gamma \equiv \eta\dot{\gamma}/\tau_y$ is the inverse Bingham number. Note that viscosity is defined as $\eta \equiv \tau/\dot{\gamma}$. We point out that $\Gamma$ can also be interpreted as a dimensionless shear rate $\Gamma = t_{\mu} \dot{\gamma}$, where $t_{\mu} \equiv \eta_{eff}/\tau_y$ is a material rearrangement timescale  \cite{coussot1995structural}. In this model, elastic deformation produces a constant $\tau = \tau_y$ at low $\dot{\gamma}$, while viscous dissipation $\eta_{eff}\dot{\gamma}$ becomes dominant for large $\dot{\gamma}$. The abrupt transition from elastic to viscous-flow states in this model corresponds to a brittle yielding behavior expected for frictional or highly annealed materials (Fig. 1c). In our earlier work, we found that the effective viscosity of the (soil + water) slurries, $\eta_{eff}$, followed the rheology of dense granular suspensions; i.e., $\eta_{eff}$ depends on the distance from jamming, $\eta_{eff} = \eta(\Delta \phi)$, where $\Delta \phi \equiv \phi_m - \phi$ and $\phi_m$ is a material-dependent jamming fraction \cite{kostynick2022rheology}. The yield stress was also found to be a function of the distance from jamming. This model cannot describe clay-rich suspensions, however, that exhibit ductile yielding (Fig. 1b). Caggioni et al. \cite{caggioni2020variations} suggested a generic constitutive relation for soft glassy materials: 
\begin{equation}
    \tau^\star = 1 + (t_{\mu} \dot{\gamma})^{1/2} + \eta\dot{\gamma}/\tau_y,
\end{equation}
where the three terms on the right hand side are elastic, plastic and viscous dissipation, respectively. For consistency with that study we follow their terminology. The term elastic describes the low shear-rate flow regime where dissipation is independent of applied shear rate \cite{bingham1922fluidity,lemaitre1994mechanics,coussot2010physical}; this regime has also been described as rate-independent plastic\cite{dimitriou2013describing}. We refer to the rate-dependent plastic regime \cite{nabizadeh2021life,jamali2019multiscale} associated with the nonlinear yielding behavior at intermediate shear rates, simply as plastic. The viscous regime at high shear rates refers to the regime where the suspension viscosity is roughly constant, corresponding to quasi-Newtonian behavior \cite{coussot1995structural}. Caggioni et al. \cite{caggioni2020variations} chose the value $n=1/2$ as the typical plastic yielding exponent for amorphous solids, and $t_{\mu}$ is a material-specific relaxation time or, equivalently, $1/t_{\mu}$ is a critical strain rate for yielding \cite{lin1999links,domenech2015rheology}. Their model was validated against experiments of various yield-stress fluids. 

While the value for the plastic yielding exponent is commonly taken to be $n=1/2$, theoretical studies indicate that $n$ may vary depending on material properties \cite{lin2018microscopic}. More broadly, if brittle materials correspond to $n=0$ and ductile materials have $n \approx 1/2$, what is the proper description for intermediate materials? For granular suspensions we posit the following: as the proportion of attractive components $\chi$ increases from $\chi = 0$ (pure granular suspensions), there is a percolation-like transition where the yielding behavior becomes progressively more ductile once stress transmission is dominated by the cohesive elements. This means that, beyond a critical value of $\chi$, the transitional (rate-dependent plastic yielding) region of the flow curve grows with increasing $\chi$ (Fig. 1c). In terms of the constitutive equation, this gradual transition from brittle to increasingly ductile behavior could be described by an increasing $n$. To allow for this possibility, we propose a generic dimensionless constitutive relation that builds on the models above:
\begin{equation}
    \tau^\star = 1+\alpha{\Gamma}^n+\Gamma.
\end{equation}
In this expression the dimensionless shear rate $\Gamma \equiv \dot{\gamma} t_{\mu}$, and the coefficient $\alpha$ describes the onset of rate-dependent plastic yielding of the suspension. For granular (sand-rich) suspensions we expect $n= \alpha = 0$, and that $t_{\mu}$ is a timescale associated with the viscous dissipation of particle motion as described by the model of Boyer et al. \cite{boyer2011unifying}. For clay-rich suspensions, however, $t_{\mu}$ should be a material relaxation time that increases with the strength of inter-particle attraction \cite{coussot1995structural} -- as observed for soft microgel pastes \cite{cloitre2003glassy}. More, clay-rich suspensions are expected to have a limiting value of $n \approx 1/2$ for ductile materials. Below we use $t_{\mu}$, $\alpha$ and $n$ as fitting parameters, and examine how they change across a transition from sand-rich (frictional) to clay-rich (cohesive) suspensions. 

\subsection*{Measuring and understanding Soft Earth suspension rheologies}
We perform rheological experiments in a stress-controlled, parallel-plate rheometer (Fig. 2a). This setup allows us to vary the gap height, and measure both shear stresses and axial forces. Following previous protocol \cite{kostynick2022rheology}, we were careful to avoid the effects of suspension wall slip, shear banding, particle sedimentation, and particle ejection at the air-suspension interface. These factors place limits on the experimentally accessible ranges of shear rates, volume fractions, and clay/sand concentrations, as discussed below.  Pre-shear protocols ensured that steady-shear flow curves were reproducible; detailed information about sample preparation and rheological measurements appears in the Methods section. We note that time-dependent changes in yield stress and viscosity -- i.e., thixotropy -- have been observed in debris-flow suspensions \cite{schippa2021thixotropic} and model mud suspensions \cite{ran2023understanding}; however, here our rheological protocol minimized these effects (see Methods section), and we focus on examining only the steady-shear behavior in order to focus on the brittle-ductile transition.  

\begin{figure}[h]
    \centering
    \includegraphics[width=1\linewidth]{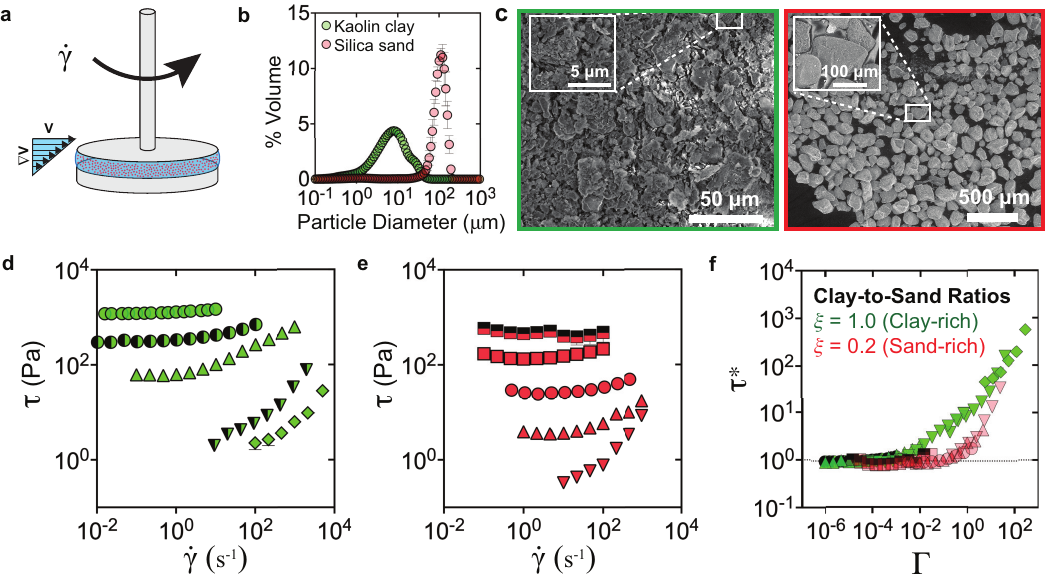}
    \caption{\textbf{Experimental Soft Earth source materials and steady-shear rheology. (a)} Parallel-plate rheology setup; arrow indicates shear direction. (\textbf{b}) Particle size distributions for kaolin clay and silica sand particles. (\textbf{c}) Scanning Electron Microscopy images of kaolin clay particles (left) and silica sand (right) particles. Insets are zoomed-in images for each case; note the stacked clay sheets for kaolin.(\textbf{d}) Steady-shear flow curves (shear stress-shear rate) of pure clay suspensions ($\xi = 1$) and (\textbf{e}) sand-rich suspensions ($\xi = 0.2$). Respective volume fractions ($\phi$) are indicated by different shapes values are indicated on the side of the plots (diamonds, $\phi$ = 0.10; invert triangle with right-half black, $\phi$ = 0.15; invert triangles, $\phi$ = 0.20; triangles, $\phi$ = 0.30; circle with left-half black, $\phi$ = 0.35; circles, $\phi$ = 0.40; squares, $\phi$ = 0.50; squares with top-half black, $\phi$ = 0.55). (\textbf{f}) Dimensionless flow curves for the data shown in (\textbf{d}) and (\textbf{e}); note similarity of the sand-rich and pure clay suspensions to California Coast Range and French Alps data, respectively (Fig. 1b).}     
\end{figure}

We prepare model Soft Earth suspensions by varying the proportions of deionized water, silica sand and kaolin clay (see Fig. 2b for particle size distributions and Fig. 2c for electron microcopy images). The total solids volume fraction (sand + clay) is denoted by $\phi$, and the relative clay fraction is defined $\xi \equiv \phi_{clay}/\phi$. Previous zeta potential ($\zeta$) measurements of the same kaolin particles in de-ionized water found $\zeta \approx -33$mV, which corresponds to moderately strong aggregation \cite{seiphoori2021tuning}. We find that below $\xi = 0.2$ and $\phi \approx 0.2$, the yield stress of the suspension is insufficient to prevent sedimentation of sand particles; this sets an experimental lower limit for our sand-rich suspensions. The upper limit of $\phi = 0.55$ for sand-rich ($\xi = 0.2$) suspensions is the point at which the sample becomes difficult to process -- i.e., it jams. For pure clay suspensions ($\xi = 1$), $\phi$ varies from a lower limit $\phi = 0.05$ to an upper achievable limit $\phi = 0.40$, beyond which the clay paste fractures. Steady-shear flow curves ($\tau$ vs. $\dot{\gamma}$) are generated for each suspension (Fig. 2d-e), following a protocol we established previously for debris-flow slurries \cite{kostynick2022rheology}.

We follow our previously documented approach \cite{kostynick2022rheology} to isolate the effect of $\phi$ on viscosity. We estimate the jamming fraction ($\phi_m$) for each suspension by fitting a divergence relation (Eq. 3 in Methods section) to the viscosity measurements, with $\phi_m$ as a fitting parameter. The yield stress $\tau_y$ for each suspension is obtained from a fit of the Herschel-Bulkley equation to the steady-shear flow curve. The observed dependence of $\tau_y$ on $\phi$ is consistent with previous observations \cite{trappe2001jamming,coussot1995structural,bougouin2022rheological} (Fig. 3a). The material timescale $t_{\mu}$ for each suspension is chosen such that all curves of differing $\phi$ values for a given value of $\xi$ collapse and yielding is centered around $\Gamma \approx 1$. We observe generally that $t_{\mu}$ decreases with increasing $\phi$ for all suspensions, but that for sand-rich suspensions ($\xi \leq 0.4$) this effect saturates (Fig. 3b). We also perform transient tests to examine yielding, where strain is applied at a constant rate $\Gamma$. In the quasi-static regime ($\Gamma \ll 1$), the stress build-up with applied strain, $\gamma$, provides a measure of elasticity, while the nature and magnitude of the stress drop at yielding indicates how brittle or ductile yielding is \cite{ozawa2018random}.

\begin{figure}[h]
    \centering
    \includegraphics[width=1\linewidth]{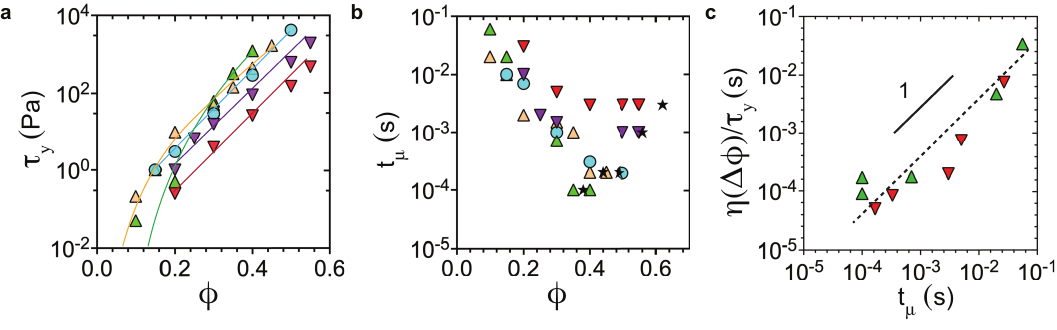}
    \caption{\textbf{Material parameters that non-dimensionalize flow curves.} The different colors correspond to the respective relative clay fraction $\xi$: 0.2 (red), 0.4 (violet), 0.6 (blue), 0.8 (peach), and 1.0 (green). For better visualization of the trends, data are classified into three distinct types: sand-rich (lower triangles; $\xi = 0.20$ and $\xi = 0.40$), clay-rich (upper triangles; $\xi = 0.8$ and $\xi = 1.0$), and intermediate ($\xi = 0.6$). \textbf{(a)} Variation in fit values of the yield stress $\tau_y$ with increasing volume fraction $\phi$, for various values of $\xi$. Trend lines for each $\xi$ are shown to guide the eye. The sand-rich mixtures (red and violet) follow a steady trend of increasing $\tau_y$ with increasing $\phi$, as reported in the literature \cite{coussot1995structural,bougouin2022rheological}. Clay-rich mixtures show a drop off in yield stress below a critical value $\phi_c$, and a critical scaling of the form $\tau_y \sim (\phi_c-\phi)^k$ which is associated with the minimal kaolin gel network required to bear the shear stress \cite{trappe2001jamming,bougouin2022rheological}.  \textbf{(b)} The change in microscopic rearrangement time ($t_\mu$) with increasing $\phi$ across $\xi$ values. The dark stars indicate the $\phi$ values (jamming distance, $\phi_m-\phi=0.03$) chosen to perform transient rheology in Fig. \ref{Fig4}b. Timescale decreases montonically with increasing $\phi$ for clay-rich suspensions, but plateaus at a constant value for sand-rich suspensions. \textbf{(c)} Empirically determined microscopic timescale ($t_{\mu}$), used to collapse the curves in Fig. 2f, scales with the estimated rearrangement timescale of a yield stress fluid ($\eta(\Delta \phi)/\tau_y$).}     
\end{figure}

Sand-rich ($\xi = 0.2$) suspensions show the same flow behavior as debris-flow slurries from the California Coast Range, which had comparable clay and sand content \cite{kostynick2022rheology}. Cohesion contributes to a yield stress at low $\Gamma$ (dimensionless shear-rate), while for large $\Gamma$ the samples behave like a granular (frictional) suspension (Fig. 2e). In other words, sand-rich samples behave as hard-particle suspensions that are weakly attractive, with essentially no intermediate ductile regime (Fig. 4a). The sand-rich suspensions for all values of $\phi$ can be collapsed onto a dimensionless Bingham curve, i.e., $\tau^\star = 1 + \Gamma$ (Figs. 2f; 5a). As expected, yield stress increases with $\phi$ (Fig. 3a), and there is no appreciable rate-dependent plastic dissipation ($\alpha = n = 0$). More, we find that $t_{\mu} \approx \eta(\Delta \phi)/\tau_y$ (Fig. 3c); that is, the experimentally determined timescale for collapsing the flow curves corresponds to the particle rearrangement timescale in a yield-stress fluid (Methods). Transient tests for nearly-jammed suspensions ($\Delta \phi \approx 0.03$) show a sharp stress drop (Fig. 4b) that is characteristic of brittle yielding. Our sand-rich suspensions jam around $\phi_m = 0.68$, consistent with typical granular materials \cite{boyer2011unifying} and the California Coast Range materials \cite{kostynick2022rheology}. For volume fractions $\phi \geq 0.40$ we observe a positive axial force (Fig. 6b), consistent with frictional effects in dense granular suspensions \cite{gamonpilas2016shear,dbouk2013normal}; the importance of this will be discussed below. Qualitatively and quantitatively, dense sand-rich suspensions exhibit brittle yielding behavior.

Pure clay ($\xi = 1$) suspensions are markedly different from sand-rich ones, and exhibit flow curves similar to the clay-rich French Alps debris-flow materials. In the dilute regime ($\phi = 0.05$) the suspension flow curve behaves like a Newtonian fluid (Fig. 2d). As $\phi$ is increased, the yield stress rapidly increases (Fig. 3a) and the material jams at roughly $\phi_m = 0.40$. Clay suspensions are known to jam at significantly lower volume fractions than granular suspensions, due to attractive particle networks that percolate across the sample \cite{bonn2017yield,coussot1995structural}. Similar to the sand-rich suspensions, for all clay suspensions with a yield stress ($\phi > 0.05$), we find that $t_{\mu} \approx \eta(\Delta \phi)/\tau_y$ (Fig. 3c; $\eta(\Delta \phi)$ estimation details in Methods section). When non-dimensionlized, the pure clay data for all $\phi$ can be collapsed onto a single curve; this curve, however, is distinct from the sand-rich data. In particular, there is the marked emergence of an intermediate yielding regime -- spanning almost three orders of magnitude in $\Gamma$ --  indicating ductile behavior (Figs. 2f; 4a; 5a). Fitting the collapsed data to Eq. 2, we find $n=1/2$ consistent with soft glassy materials, and $\alpha \approx 10$ (Fig. 5c). We suspect that the exponent $n \approx 1/2$ is related to the mechanism of plastic dissipation, while the coefficient $\alpha$ is determined by the strength of inter-particle attraction. To test this idea, we conduct additional experiments with identical clay and water content, but varying ionic strength through the addition of 0.001 M and 0.1 M NaCl. Salt screens the repulsive net charge of clay particles and facilitates stronger aggregation \cite{seiphoori2021tuning,israelachvili2022surface}. As expected, we find that $\alpha$ increases systematically with ionic strength, but $n$ remains constant (Fig. 5b,c; Supplementary Information Fig. S2 and Fig. S3). Quasi-static strain tests of nearly-jammed clay-rich suspensions ($\Delta \phi \approx 0.03$; corresponding $\phi$ values marked in Fig. 3b) show a very small stress drop after yielding (Fig. 4b), and negligible axial force (Fig. 6b). In sum, clay suspensions exhibit all of the hallmarks of ductile yielding, and do not show any effects of frictional interactions between particles. Our results support the characterization of clay suspensions as a gel.

\begin{figure}[h]
    \centering
    \includegraphics[width=1\linewidth]{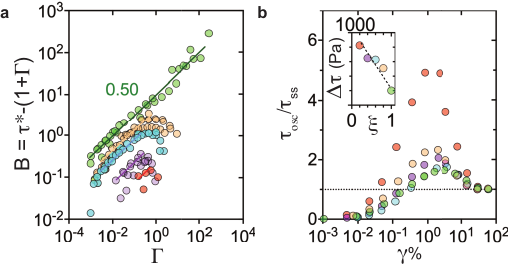}
    \caption{\textbf{Signatures of the brittle-ductile transition. (a)} The Bingham part of the constitutive equation $\tau^\star-1-\Gamma$ plotted against the dimensionless shear rate. The curves shown correspond to the plastic term $\alpha\Gamma^n$ in the constitutive equation 2. Colors correspond to $\xi$ values in Figure 3. Sand-rich suspensions have almost no plastic regime. With increasing clay content $\xi$, we see the growth of a plastic regime and the emergence of a robust exponent $n=1/2$. (\textbf{b}) Stress-strain curves for suspensions at all values of $\xi$, where the oscillatory stress $\tau_{osc}$ is normalized by the steady-state stress after yielding $\tau_{ss}$, and strain is in units of percent. Inset shows that the stress overshoot systematically decreases as $\xi$ increases; this behavior is characteristic of a brittle to ductile transition (compare to Fig. 1 in ref. \cite{ozawa2018random}).}     
\end{figure}

\subsection*{Rheological signatures accompanying brittle-to-ductile transition}

We can now examine the transition from brittle to ductile behavior as relative clay content is systematically changed between $\xi = 0.2$ and $\xi = 1$. Transient shear tests show that as $\xi$ increases, the stress drop associated with yielding systematically decreases (Fig. 4b inset). For steady-shear flow curves, we observe that the failure envelope -- defined by the range in $\Gamma$ associated with the transitional regime between elastic (rate-independent) and viscous flow -- increases systematically with increasing $\xi$ (Figs. 4a, 5a, 6a). One way to examine this failure envelope is to plot the Bingham part of dimensionless flow curves, $B = \tau^{*} - 1 - \Gamma$, against the dimensionless shear rate $\Gamma$. If the data follow a (dimensionless) Bingham relation, the resulting curve would be flat with all values nearly zero. Indeed, we observe that sand-rich suspensions exhibit almost no plastic regime. The data reveal that, as $\xi$ increases, there is a systematic growth of the plastic regime that is characterized by the emergence of a robust $n=1/2$ scaling region (Fig. 4a). Fitting Eq. 2 to the data, we observe that $n$ increases with $\xi$ from a lower limit of $n=0$ at $\xi = 0.20$ to a maximum of $n=1/2$ at $\xi = 1$ (Fig. 5b). We also observe that $\alpha$ increases systematically with clay concentration and ionic strength (Fig. 5c), consistent with our proposal that this parameter is related to the strength of cohesive bonds.  The dependence of $\alpha$ on $\xi$ is nonlinear, with a marked increase occurring in the vicinity of $\xi \approx 0.5$ (Fig. 5c). Indeed, this is where $\alpha$ becomes order one, meaning that the plastic stress term is comparable in scale to the elastic stress term. Previous studies have mapped gelation in colloid and polymer systems to percolation \cite{de1976relation, del1999elastic, stauffer2005gelation}. If the percolation interpretation is relevant for our system, it implies that increasing $\xi$ corresponds to increasing the cluster size of the attractive clay network -- and that the brittle-ductile transition occurs when these cohesive particle networks percolate across the sample. Axial forces become negligible for $\xi> 0.6$, suggesting that frictional granular contacts are unable to percolate the sample for larger clay concentrations (Fig. 6b). A different take on criticality is to examine the behavior of the ``yielding exponent'', defined in elasto-plastic models as $\beta \equiv 1/n$ \cite{lin2018microscopic}. In this view, $\beta = 2$ in the fully ductile regime corresponding to pure clay suspensions ($\xi = 1$). As $\xi$ decreases toward a critical value $\xi_c$ associated with sand-rich suspensions, $\beta$ diverges as yielding becomes increasingly discontinuous (Fig. 5b inset). In the limit that $\beta \rightarrow \infty$, which corresponds to roughly $\xi_c \approx 0.2$ for our suspensions, suspension yielding is discontinuous and reduces to classic Coulomb failure. 

\begin{figure}[h]
    \centering
    \includegraphics[width=1\linewidth]{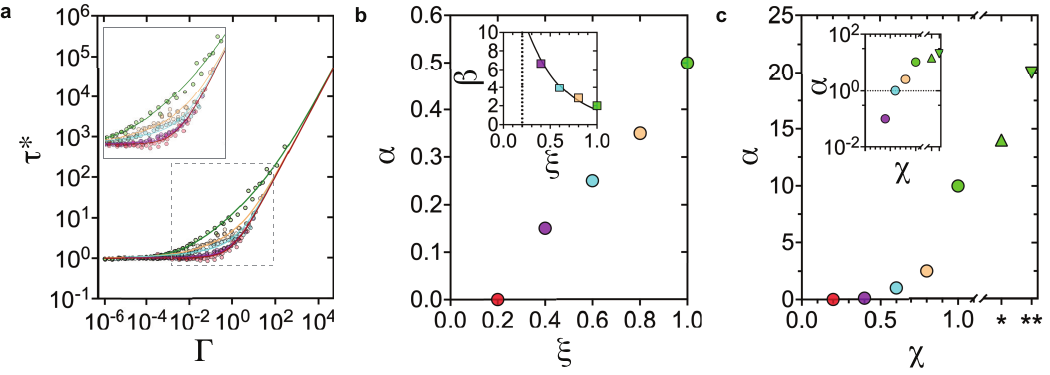}
    \caption{\textbf{Constitutive relations for Soft Earth suspensions. (a)} Non-dimensionalized flow curves of the form $\tau^\star = 1+\alpha\Gamma^n+\Gamma$ for all values of $\xi$. Note changes in the vicinity of yielding (inset), corresponding to increasing envelope of rate-dependent plasticity with increasing $\xi$. Colors correspond to sand-clay mixtures as indicated in Fig. 3. (\textbf{b}) The exponent $n$ in the constitutive equation increases gradually from a minimum $n=0$ at $\xi = 0.20$ associated with brittle failure, to a maximum $n = 1/2$ at $\xi = 1$ corresponding to fully ductile failure. Increasing the attraction strength of clay did not change $n$. Inset: the yielding exponent $\beta = 1/n$ diverges as $\xi \rightarrow \xi_c$ from above, where the critical relative clay fraction $\xi_c = 0.2$. This is another way to show the continuous nature of the ductile to brittle transition in Soft Earth suspensions. (\textbf{c}) The fit parameter $\alpha$, the coefficient of the rate-dependent plastic term in equation 2, plotted against the increasing attractive/cohesive elements ($\chi = \chi(\xi,\zeta)$). The prefactor $\alpha$ increases with both clay concentration and ionic strength, indicating that it reflects the strength of inter-particle attraction. The $\star$ denotes clay suspension with $10^{-3}$ M NaCl (upper triangle; $\xi = 1.0$) and $\star\star$ denotes the sample with $10^{-1}$ M NaCl (lower triangle; $\xi = 1.0$). Inset represents the same plot, with the y-axis in logarithmic space. Above the dotted line ($\alpha > 1$) the plastic term becomes important.}     
\end{figure}

It is worth pointing out that fixing $n=1/2$ in Eq. 2, rather than allowing it to vary, results in reasonably good fits to the data as well (Supplementary Information Fig. S4). Given that this model has only one free parameter ($\alpha$) rather than two, from a statistical point of view the fixed $n=1/2$ relation would seem to be preferable. There are two reasons, however, to explore a variable $n$ model. First is to examine whether a robust value for $n$ emerges with increasing clay content (decreasing sand content), rather than imposing it. Indeed, it appears that increasing clay fraction results in an increasing envelope of plasticity characterized by $n=1/2$ scaling (Fig. 4a). From a physical point of view this suggests that there may be some kind of universality in the plastic exponent $n=1/2$. On the other hand, allowing $n$ to vary accounts for transitions into and out of the plastic regime. This leads to a second, practical reason to allow $n$ to vary: the fixed $n=1/2$ model systematically overfits the flow curves in the viscous (large $\Gamma$) regime, while the variable $n$ model does a better job (Supplementary Information Fig. S4).

Additionally, it is important to point out that the rheology data of Soft Earth suspension mixtures agree with the earlier observations of the viscosity enhancement in yield stress fluids with the addition of non-colloidal particles \cite{mahaut2008yield,dagois2015rheology}. For example, consider two samples at constant $\phi_{clay} \approx 0.11$ and varying $\phi_{sand}$: (i) $\phi_{sand}=0$ $(\xi = 1.0)$ and (ii) $\phi_{sand}=0.18$ $(\xi = 0.4)$. The respective steady shear curves (Fig. 2d, Supplementary Information Fig. S3a) show the viscous dissipation in $\phi_{sand}=0.18$ is higher than pure clay mixture, at any given $\dot{\gamma}$.

\begin{figure}[h]
    \centering
    \includegraphics[width=1\linewidth]{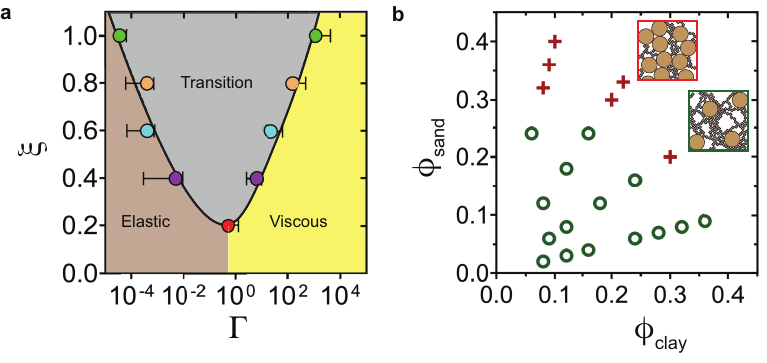}
    \caption{\textbf{The phase space of Soft Earth suspension behaviors. (a)} Experimentally determined phase space of elastic (rate-independent plastic), viscous and transitional (rate-dependent plastic) dissipation regimes; compare to the proposed flow phase space in Fig. 1c. Colors correspond to $\xi$ values in Fig. 3. Error bars indicate SD. (\textbf{b}) Phase space of axial force for Soft Earth suspensions containing different mixtures of sand and clay. Phase space of axial forces for Soft Earth suspension mixtures as a function of constituent clay and sand content. The data is shown in binary mode; circles denote no measurable axial force, while crosses denote positive measurable axial force. The positive axial force is speculated to be associated with formation of load bearing force chains in sand-rich dense suspensions -- i.e., granular-frictional effects (red box). Only suspensions with relatively high sand content ($\phi_{sand}$) exhibit a positive axial force; as clay content ($\phi_{clay}$) increases beyond a critical value, the axial force becomes negligible as cohesion dominates over friction (green box). Inset boxes correspond to the posited clay-rich and sand-rich load-bearing microstructures in Fig. 1b.}     
\end{figure}

\section*{Discussions}
The origins of complex flow behaviors for Soft Earth suspensions arise from two critical points. First is the well known jamming transition, in which viscosity diverges with increasing volume fraction \cite{boyer2011unifying, kostynick2022rheology,pradeep2021jamming}, as $\eta(\phi) \sim (\Delta \phi)^{-2}$. The second, revealed here, is the ductile to brittle transition wherein $\beta \rightarrow \infty$ as $\xi \rightarrow \xi_c$.
Sand-rich slurries match the behavior of granular debris flows; they exhibit brittle yielding corresponding to a rapid transition from frictional to viscous stress dissipation. These mixtures also exhibit positive axial forces at high volume fractions, similar to sheared granular suspensions \cite{dbouk2013normal,gamonpilas2016shear}. From a microstructural perspective, we speculate that the positive axial forces are due to the stress-bearing force chains arising from frictional (sand) contact networks, similar to experimental shear thickening suspensions \cite{hsu2018roughness,pradeep2021jamming,royer2016rheological}. In free-surface flows, these granular effects can give rise to dilatancy and transient pore-pressure dynamics that must be considered for simulating debris flows \cite{iverson2014depth}. Even suspensions with moderate clay content ($\xi = 0.20$) still behave as granular suspensions; however, the yield stress arising from cohesion must be taken into account \cite{kostynick2022rheology}. Clay-rich slurries reproduce the observed rheology of Cohesive debris flows; ductile yielding arises due to strong rate-dependent plastic stress dissipation. Strong cohesion may act as a confining pressure \cite{vo2020additive} that shuts off axial forces; these suspensions can be modeled as gels, where dilatancy and related effects are negligible 6b). Two important concepts emerging from granular and suspension rheology are used to build the constitutive Eq. 2. First, there are elastic, plastic and viscous stresses that arise from the distinct material components -- frictional sand, attractive clays, and water -- and these stresses may be additive \cite{vo2020additive, caggioni2020variations, tapia2022viscous}. Second, rheologies can be assembled by nondimensionalizing the shear rate with an appropriate microscopic rearrangement timescale \cite{boyer2011unifying}. In contrast to (cohesionless) granular suspensions, however, the yield stress -- rather than the confining pressure -- is the relevant quantity. Thus, the microscopic timescale $t_{\mu} \approx \eta(\Delta \phi)/\tau_y$. Field-scale debris flows may introduce inertial effects for both particles (collisions) and fluid (turbulence) \cite{kostynick2022rheology} that are not considered here. Exploring the inertial regime for cohesive suspensions -- as has been done recently for granular suspensions \cite{tapia2022viscous} -- is a logical next step.  

The brittle-ductile transition in our suspensions appears to be due to a handoff from frictional to cohesive control, when clay aggregates percolate the sample. This transition is manifest as a continuous increase in plasticity, and decrease in granular effects such as a axial force, as the relative clay concentration for dense suspensions is increased (Figs. 4-6). The exponent $n=1/2$ emerges as a robust scaling for the plastic yielding regime in our suspensions, and is consistent with a variety of non-frictional yield-stress fluids such as emulsions and colloidal suspensions that were examined by Caggioni et al. \cite{caggioni2020variations}. Transitions into and out of the rate-dependent regime, however, vary as a function of $\xi$. This variation is not captured by a fixed $n=1/2$ model, which is why allowing $n$ to vary provides a better fit to the data. Although there are no modeling studies to directly compare to our observations, some of our findings could be anticipated from simulations of idealized amorphous solids, which helps to understand the origins of this transition. Granular simulations found that frictional particles exhibit brittle yielding, while the same particles without friction showed ductile failure \cite{karimi2019plastic}. A different simulation study of an idealized glass found that increasing the lengthscale of inter-particle interactions was sufficient to drive a transition from brittle to ductile failure \cite{dauchot2011athermal}. Our brittle and ductile suspensions exhibit similar behavior to highly and poorly annealed glasses, respectively \cite{ozawa2018random}. These results suggest that adding clay may lubricate contacts among sand grains \cite{ancey2007plasticity}, increase the lengthscale of cooperative particle motion \cite{bonn2017yield}, and/or maintain ``soft spots'' that do no anneal under shear \cite{ozawa2018random}. Although these mechanisms are different, they all act to delocalize failure and prevent formation of a slip plane -- which is the origin of brittle failure. Probing the brittle-ductile transition in an optically-transparent cohesive-frictional mixture, such as clear particles suspended in a gel \cite{kammer2022homogenization}, would allow visualization of the microscopic dynamics underlying failure; we leave this for future work.

We conclude with considering the consequences of our findings for yielding behavior in the Soft Earth. Sudden liquefaction of soil, due to seismic shaking or rainfall, is typically limited to water-saturated sand-rich materials that exhibit brittle failure due to pore-pressure induced dilation \cite{iverson2014depth, huang2017hazard}. Cohesive soils tend to form slow ``earthflows'' that creep ductily \cite{evans2001landslides}. Shale, a type of sedimentary rock, has been shown to undergo a transition from brittle to ductile failure when the fraction of cohesive materials (clay + organics) exceeds 0.35 \cite{wang2019effect}, consistent with our data and a percolation-like transition. Finally, deep in the Earth's lithosphere the rock becomes soft, and the accompanying brittle-to-ductile transition exerts a strong control on the source, rupture dynamics and magnitude of earthquakes \cite{kato2021generation}. While these geophysical problems involve disparate materials, scales and stresses, the discovered connections -- from idealized amorphous solids to our model Soft Earth suspensions to natural debris-flow materials -- embolden us to seek commonality in failure across more Earth materials.

\section*{Methods}

\subsection*{Soft Earth suspension preparation}
 Silica sand and kaolin clay (semi-dry and air floated) were purchased from AGSCO Corporation (Pine Brook, NJ, USA) and Unimin Corporation (McIntyre, GA, USA), respectively. Both model clay and sand particles were used as obtained without further purification. Particle size distributions were measured using a Beckman-Coulter Particle Size Analyzer LS13-320. Grain size was determined in 114 log-spaced bins over the range 0.04 $\mu$m to 2000 $\mu$m. The particle size distributions in both kaolin clay and silica sand contained a single mode, and thus had distinct average particle size population. Microscopic images of silica sand and kaolin clay were obtained using Scanning Electron Microscope (FEI Quanta 600 Environmental Scanning Electron Microscope). The images were acquired at an accelerating voltage of 15 kV and a water vapor pressure of 0.75 torr. 

Model Soft Earth suspensions were prepared at desired final concentrations ($\phi$) by mixing the three components -- sand, clay, and water -- in varying proportions. First, a pre-determined amount of kaolin clay was mixed with de-ionized water for 30 mins in a high-shear mixer. The mixture was left overnight to allow the clay particles to absorb water and reach an equilibrium state. In the second step, silica particles were added to the clay suspension and further mixed in the high-shear mixer for $\sim$10 mins, to ensure uniform distribution. For sample preparation purposes we use density of particles (both sand and clay) and deionized water as 2650 kg/m$^3$ and 1000 kg/m$^3$, respectively.

\subsection*{Rheological characterization}
All rheological measurements are carried out using a TA Instruments DHR-3 model rheometer, with advanced strain and stress control, using a 40mm parallel-plate setup at 25$^0$C. To reduce the sample slip effects at the boundaries during our measurements, both the top and bottom plates were modified by attaching 50-Grit size serrated surfaces. This is equivalent to roughness length scale $\approx$ 300 $\mu$m, which is three times the largest particle size. We maintained a gap height of 1mm for all the experiments ($\sim$ 10 times the largest particle size), which is shown to reduce particle confinement effects during the shear flow \cite{peyla2011new}. All the experiments performed are at constant volume, and thus our system is an NVT ensemble.

All suspensions were pre-sheared to ensure that the samples did not have shear history, allowing us to generate reproducible steady-shear and oscillatory rheological measurements. In the first step, samples were sheared using a large-amplitude oscillatory protocol ($\gamma$ = 500\%, $\omega$ = 10 rad/s) to minimize the directional bias in the microstructures that come from steady shear protocols \cite{choi2020optimal}. The destruction of the internal structure was monitored through changes in the suspension elastic modulus ($G'$), which decreased and then plateaued within 60s, for all our samples. During the second step, the sample was allowed to recover into a reproducible average microstructure. The recovery time was estimated using a small-amplitude oscillatory protocol ($\gamma$ = 0.05\%, $\omega$ = 1 rad/s), which resembles near-equilibrium rheology measurements and the linear deformation regime of the material \cite{pradeep2022hydrodynamic}. We monitored the value of the elastic modulus, which grew as the suspension structure recovered, and plateaued to a constant value for all our samples within 30-80s (Supplementary Information Fig. S1).

Following the pre-shear, we follow a robust protocol to reduce the thixotropic effects \cite{schippa2021thixotropic} and associated rheological flow hysteresis \cite{ran2023understanding} in our rheological measurements. After pre-shearing, we perform a downsweep: each sample was sheared from the high shear rate point ($\dot\gamma_{max}$) to low shear rate point ($\dot\gamma_{min}$), dwelling at each point for 100s to achieve a stable value for $\tau$. Downward sweep is followed by an upward sweep from $\dot\gamma_{min}$ to $\dot\gamma_{max}$, with stress equilibration at each $\dot\gamma$. The shear stress values reported are averaged values of upsweep and downsweep, and are thus devoid of first-order flow hysteresis effects \cite{puisto2015dynamic}. The maximum and minimum shear stress limits in our flow curves were determined by the interfacial and gravitational stresses, respectively, set by the length scale of the largest particles in our system. The interfacial stress is estimated as  $\tau_{max} \sim \gamma_{a-w}/a$, where $\gamma_{a-w}$ is the interfacial tension at the air-water interface and $a$ is the average radius ($\approx$ 50 $\mu$m) for the larger particles (silica sand). This value sets the shear stress beyond which particles may eject out of the fluid-air interface. The lower shear stress limit is set by the gravitational force acting at the particle scale given by $\tau_{min} \sim \Delta\rho ga$, where $\Delta\rho$ is the density difference between particle and the suspending fluid ($\approx$1650 kg/m$^3$) and $g$ is acceleration due to gravity (9.8 m/s$^2$). Below this stress, the sedimentation effects due to settling of silica grains is important \cite{kostynick2022rheology}. Transient rheological characterizations reported in Fig. 4b were performed using amplitude sweep protocol at constant oscillatory rate ($\omega = 1$ rad/s). Samples prepared close to their respective jamming points ($\Delta \phi \approx 0.03$) were loaded on the rheometer, followed by amplitude sweep from 0.01 - 100 \% strain units, to generate the shear stress-strain ($\tau_{osc}-\gamma \%$) curves.

We interpret the sign of the axial force exerted by the suspension on the top rheometer geometry. In a constant volume plate-plate setup, normal stress difference is generally estimated from the axial thrust recorded by the transducer \cite{macosko_1994}. For such a calculation, it is assumed that the normal stress in the radial direction is balanced by the difference between the atmospheric pressure (that holds the suspension boundaries in place) and interfacial stress \cite{hsiao2019experimental}. However, this analysis ignores particle protrusion effects, which were observed in dense granular suspensions \cite{brown2012role}. Therefore, we interpret the raw axial force data for all our suspension mixtures qualitatively, which is shown in Fig. 6b.

\subsection*{Estimating microscopic timescales}
We implement a weight-averaged method, using the clay ratio ($\xi$), to theoretically determine the microscopic rearrangement timescale in our Soft Earth suspension mixtures across the frictional to cohesive regimes. The theoretical particle rearrangement timescale in yield stress fluids is estimated as a ratio of an effective suspension viscous dissipation and the yield stress, $\eta_{eff}/\tau_y$. Here, $\tau_y$ is the suspension yield stress, which dominates the low shear-rate behavior. The term $\eta_{eff}$ is the suspension viscosity in the high shear-rate limit, where the microstructure is broken down to clusters or particles that dissipate \emph{via} hydrodynamic or frictional interactions, respectively. This results in a quasi-Newtonian viscous behavior at high shear that diverges on approach to the jamming point, $\phi \rightarrow \phi_m$, where ($\phi_m$) is a characteristic property of the material. Therefore, the viscous dissipation at high shear can be expressed as a function of the jamming distance, $\eta_{eff} = \eta(\Delta\phi) \sim (\phi_m - \phi)^{-2}$, as done before \cite{pradeep2021jamming,kostynick2022rheology}. This allows us to fit $\phi_m$, and estimate $\eta(\Delta\phi)$ and $t_\mu$ for each suspension mixture. For granular suspensions $\eta(\Delta\phi)$ is estimated as \cite{boyer2011unifying}:
\begin{equation}
    \left(\frac{\eta(\Delta\phi)}{\eta_s}\right)_{sand}=1+\frac{5}{2}\phi\left(\frac{\phi_m}{\phi_m-\phi}\right)+\mu_\xi\left(\frac{\phi}{\phi_m-\phi}\right)^{2}.
\end{equation}

The terms $\phi_m$ and $\mu_\xi$ are fitting parameters, that depend on constituent material properties and the clay fraction $\xi$ (Supplementary Information Fig. S5). We model the high-shear dissipation for cohesive clay suspensions by assuming that viscosity divergence is governed by intercluster hydrodynamics, as has been done for gels \cite{domenech2015rheology},:
\begin{equation}
  \left(\frac{\eta(\Delta\phi)}{\eta_s}\right)_{clay}=\eta_s\psi(\phi).
\end{equation}
Here, $\psi(\phi)$ is a hydrodynamic function that takes the form of the Kreiger-Dougherty viscosity divergence, $\psi(\phi)=(1-\phi/\phi_m)^{-2}$, where $\phi_m$ is a fitting parameter. Assuming additivity of stresses \cite{vo2020additive}, we estimate the viscous dissipation $\eta(\Delta\phi)$ for a given $\xi$ using a weighted-average method:
\begin{equation}
    \eta(\Delta\phi)=\eta(\Delta\phi)_{sand}(1-\xi)+\eta(\Delta\phi)_{clay}(\xi).
\end{equation}
We fit the values of $\phi_m$ and $\mu_\xi$ (Supplementary Information Fig. S5) to estimate viscosity as a function of jamming distance and, finally, to determine theoretical microscale rearrangement timescale. The fit jamming point ($\phi_m$) decreases with decreasing sand concentration, and plateaus at $\approx 0.40$ at $\xi = 1.0$. The theoretical estimates scale linearly with the empirical microscopic timescales $t_\mu$ (Fig. 3c, Supplementary Information Fig. S2 and S3).

\bibliography{sn-bibliography}

\section*{Declarations}

\bmhead{Acknowledgments} We thank Jamie Ford (Nanoscale Characterization Facility, Singh Center for Nanotechnology) and John Ruck (University of Pennsylvania) for their help with particle characterizations; and Robert Kostynick (Washington University in St. Louis), Alban Sauret, Eckart Meiburg, Tom Dunne (University of California, Santa Barbara), Alban Th\'ery, and Pedro Ponte Castañeda (University of Pennsylvania) for their inspiration and feedback. The research was supported by National Science Foundation (NSF) Materials Research Science and Engineering Center (NSF-DMR-1720530) and Army Research Office (ARO Grant W911NF2010113) grants to D.J.J. and P.E.A., and was supported in part by grant NSF PHY-1748958 to the Kavli Institute for Theoretical Physics (KITP).

\bmhead{Author Contributions} S.P. led all aspects of this study: the conceptual framework, experimental design and execution, and analysis of results. P.E.A. and D.J.J. were co-equal supervisors of the work, and participated in data analysis and interpretation. The manuscript was written collaboratively by all authors.

\bmhead{Competing Interests} Authors declare no competing interests.  

\newpage

\pagestyle{empty}
\renewcommand{\thefigure}{S\arabic{figure}}
\setcounter{figure}{0}

\section*{Supplementary Information}

\begin{figure}[H]
    \centering
    \includegraphics[width=0.60\linewidth]{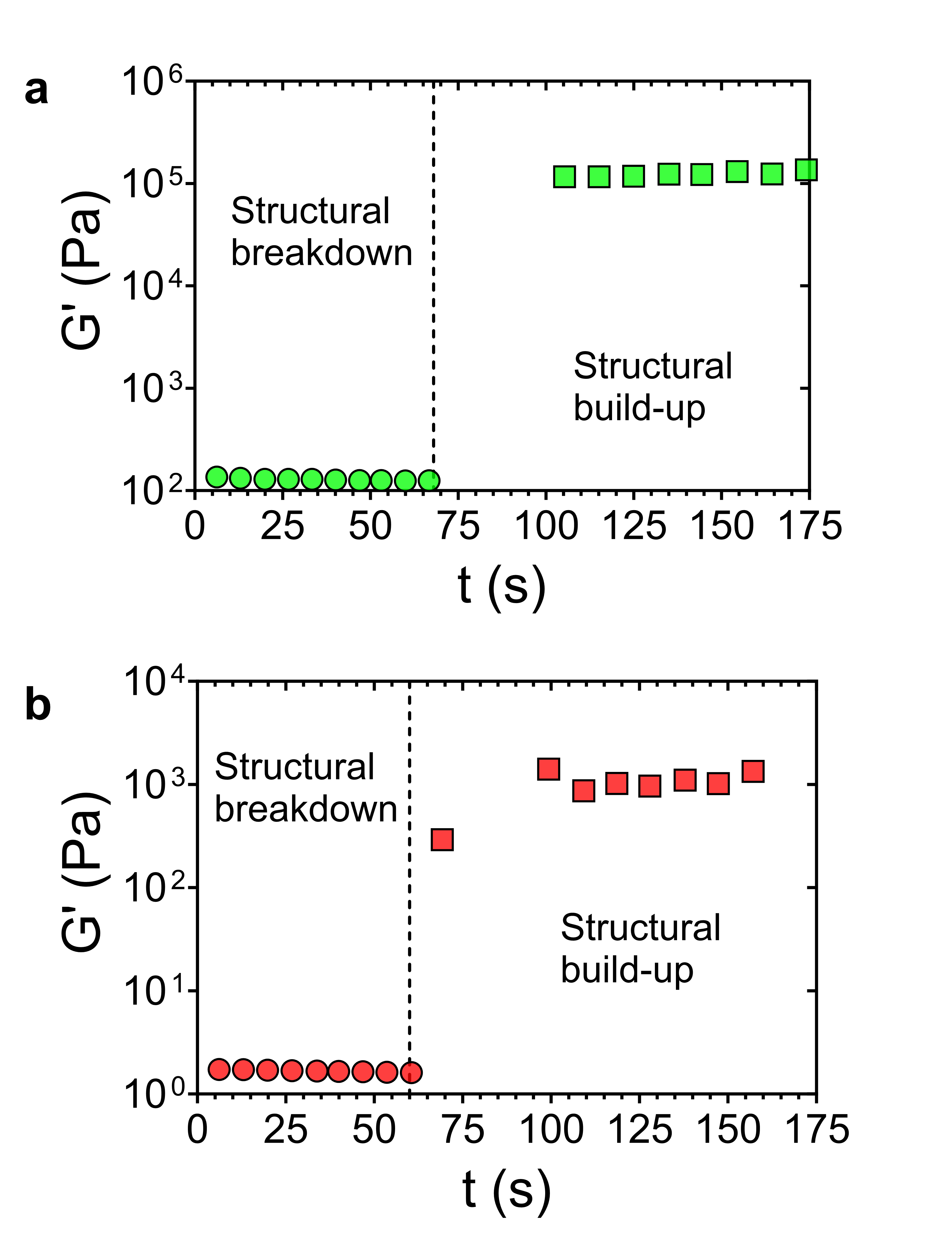}
    \caption{Sample preshear protocol for \textbf{(a)} pure clay ($\xi = 1.0$; green) and \textbf{(b)} sand-rich ($\xi = 0.2$; red) suspensions at $\phi = 0.4$. In both cases, a combination of high strain amplitude (500\%) and oscillation rate (10 rad/s) destroys the structure and associated residual stress. This results in low shear modulus $G'$ values (circles). During the second step (beyond vertical dashed lines), the microstructure is allowed to recover by monitoring increase in $G'$ (squares) by performing oscillations in the linear limit (0.01\%, 1 rad/s) of the Soft Earth suspensions. All samples were subjected to similar preshear protocol before steady shear experiments.}\label{Fig S1}     
\end{figure}

\begin{figure}[H]
    \centering
    \includegraphics[width=0.75\linewidth]{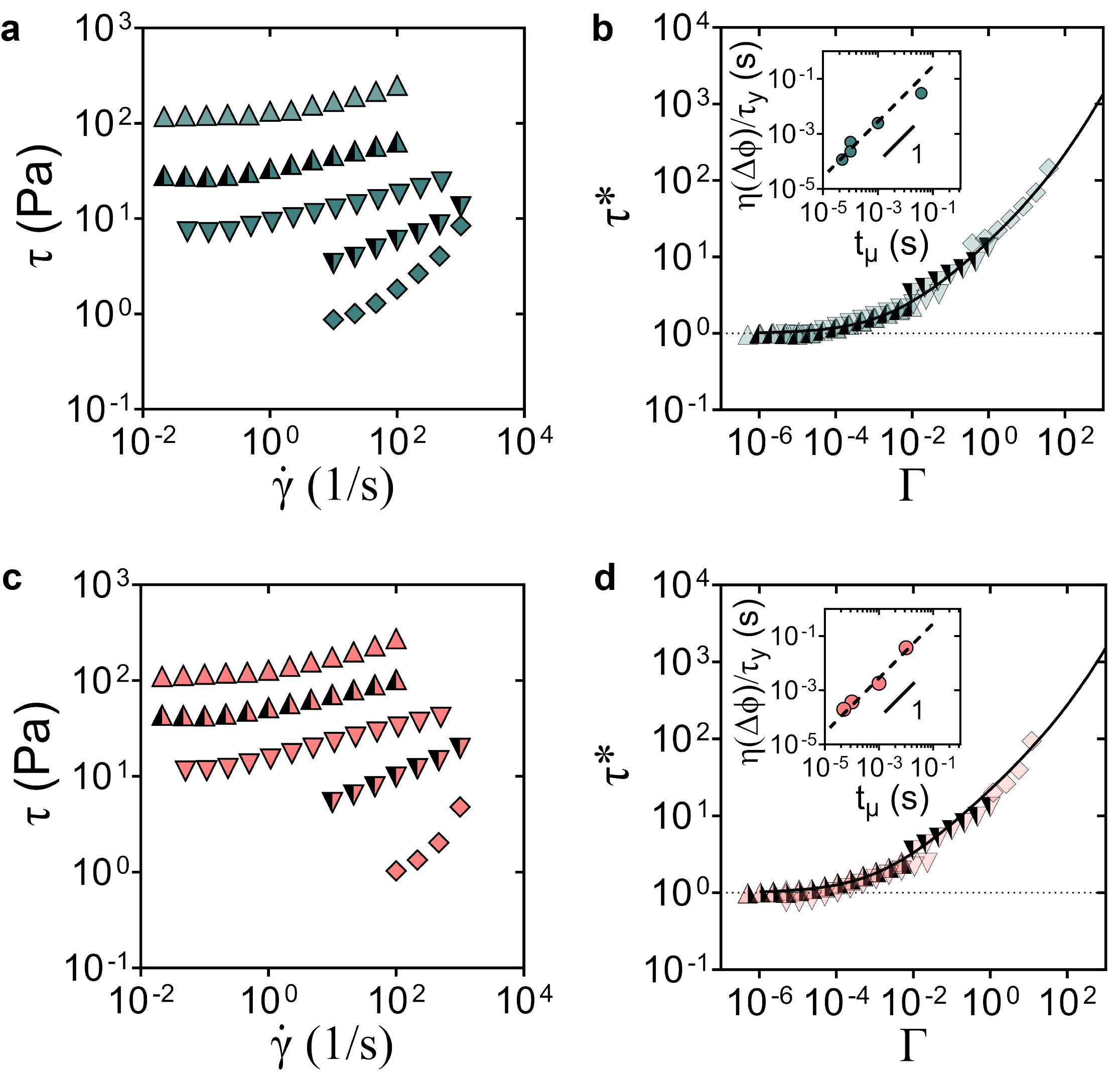}
    \caption{Steady shear rheological flow curves of kaolin clay ($\xi = 1$) with (\textbf{a}) 0.001 M and (\textbf{c}) 0.1 M NaCl. Respective volume fractions ($\phi$) are indicated by different shapes (diamonds, $\phi$ = 0.10; invert triangles with right-half black, $\phi$ = 0.15; invert triangles, $\phi$ = 0.20; triangles with right-half black, $\phi$ = 0.25; triangles, $\phi$ = 0.30). The non-dimensionalized flow curves ($\tau^\star - \Gamma$) are shown in plots (\textbf{b, d}), respectively. Corresponding linear correlations between the empirical rearrangement time scales ($t_\mu$) used to collapse the steady shear curves and the theoretical microscopic rearrangement time scales ($\eta(\Delta\phi)/\tau_y$; refer ``Methods" section in the main text for details) are shown in the inset of the collapsed plots \textbf{b} and \textbf{d}.}\label{Fig S2}     
\end{figure}

\begin{figure}[]
    \centering
    \includegraphics[width=0.75\linewidth]{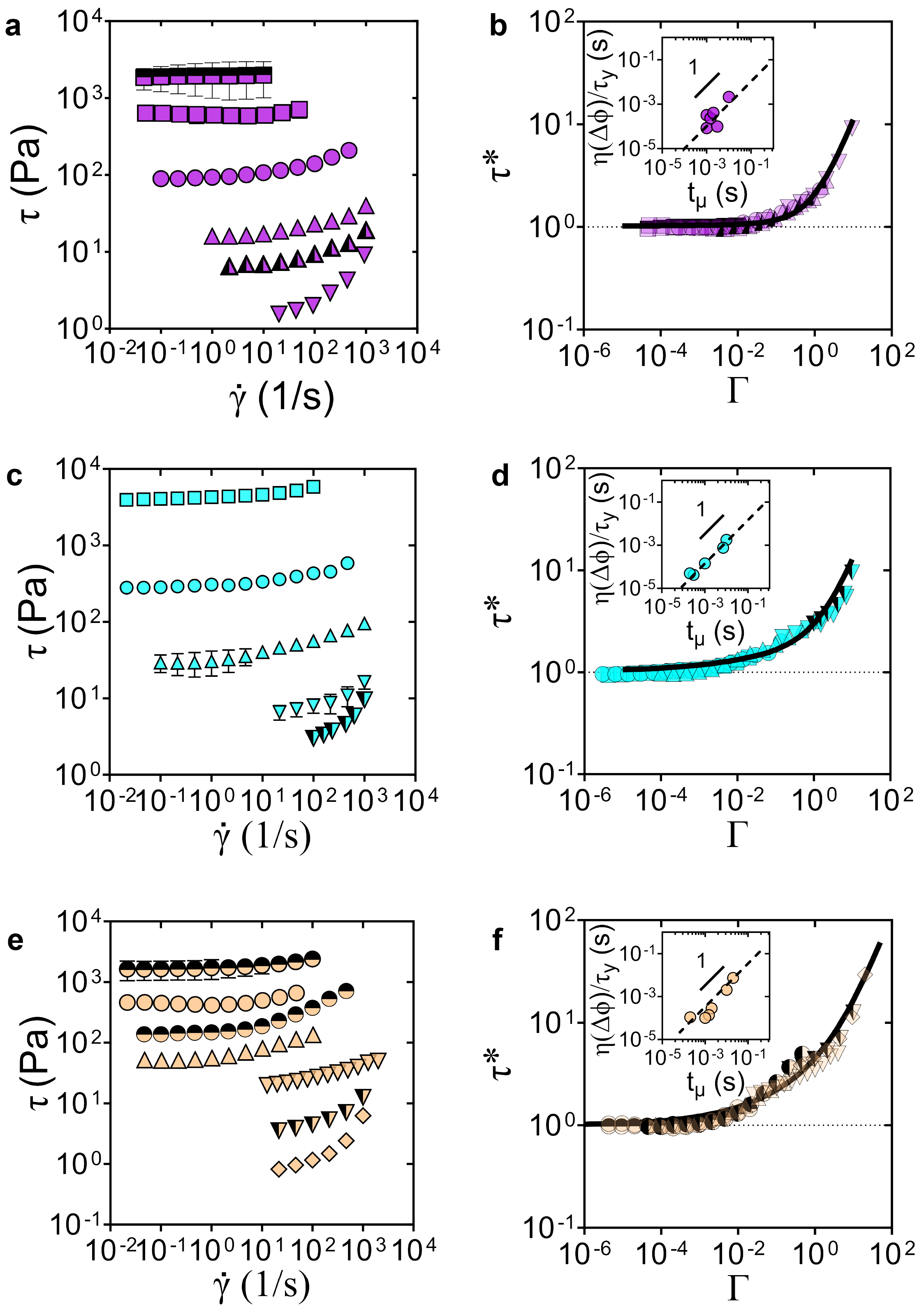}
    \caption{Steady shear rheological flow curves of sand-clay suspension mixtures at (\textbf{a}) $\xi = 0.4$, (\textbf{c}) $\xi = 0.6$, and (\textbf{e}) $\xi = 0.8$. Respective volume fractions ($\phi$) are indicated by different shapes (diamonds, $\phi$ = 0.10; invert triangles with right-half black, $\phi$ = 0.15; invert triangles, $\phi$ = 0.20; triangles with right-half black, $\phi$ = 0.25; triangles, $\phi$ = 0.30; circle with left-half black, $\phi$ = 0.35; circles, $\phi$ = 0.40; circles with top-half black, $\phi$ = 0.55; squares, $\phi$ = 0.50; squares with top-half black, $\phi$ = 0.55). The non-dimensionalized flow curves ($\tau^\star - \Gamma$) are shown in plots (\textbf{b, d, f}), respectively.Corresponding linear correlations between the empirical rearrangement time scales ($t_\mu$) used to collapse the steady shear curves and the theoretical microscopic rearrangement time scales ($\eta(\Delta\phi)/\tau_y$; refer ``Methods" section in the main text for details) are shown in the inset of the collapsed plots \textbf{b}, \textbf{d}, and \textbf{f}.}\label{Fig S3}
\end{figure}

\pagestyle{empty}

\begin{figure}[h]
    \centering
    \includegraphics[width=1.1\linewidth]{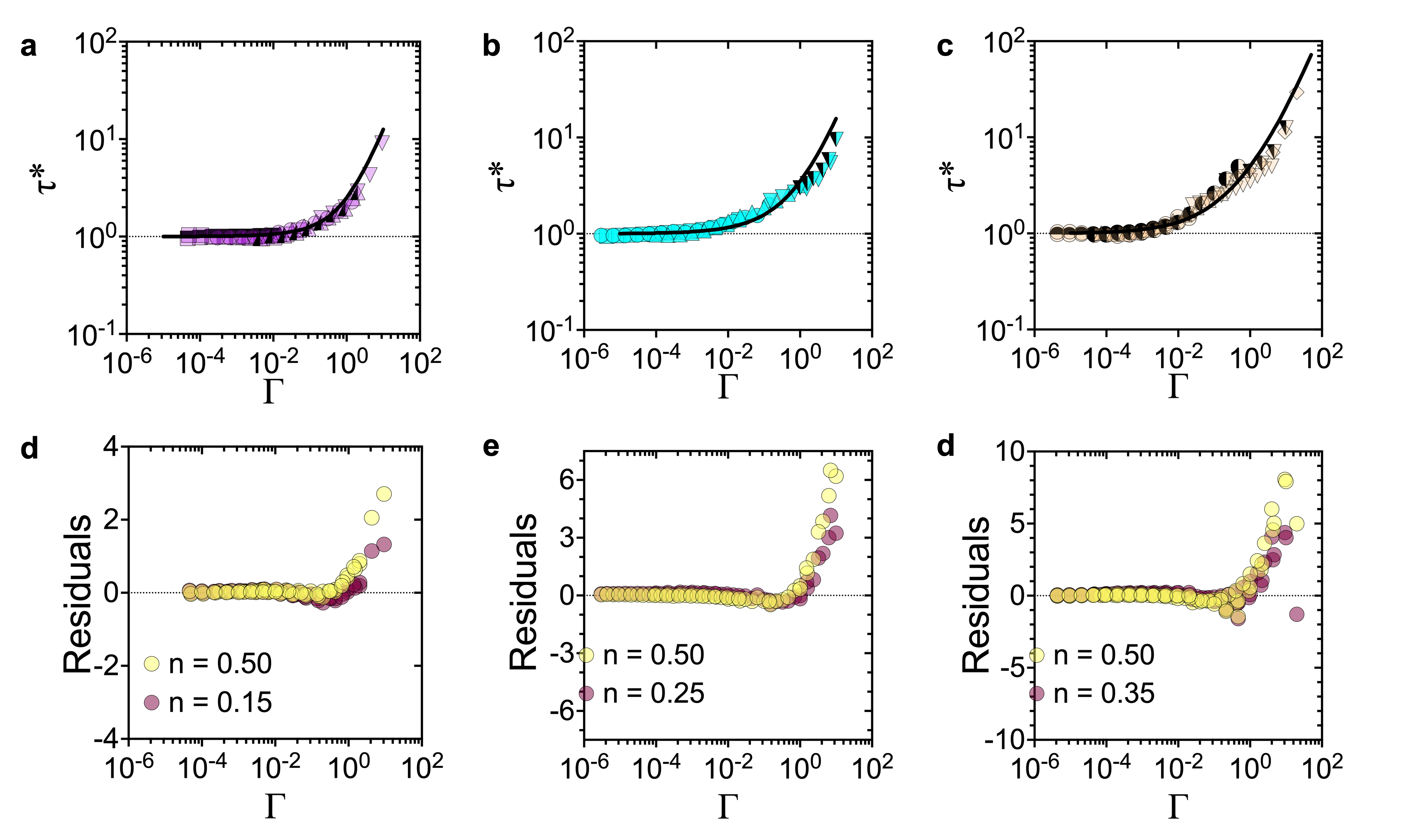}
    \caption{The fit parameters of the rheological constitutive equation: $\tau^\star = 1+\alpha\Gamma^n+\Gamma$, plotted against the increasing attractive/cohesive elements ($\chi = \chi(\xi,\zeta)$) in the Soft Earth suspension mixture. (\textbf{a})) The plastic prefactor $\alpha$, continuously increases with increasing attractive components in the system. (\textbf{b}) The shear thinning exponent $n$, increases with proliferating cohesive elements in the suspension mixture. Unlike $\alpha$, the value of $n$ saturates at 1/2 (dotted lines in \textbf{b}) in the pure clay suspension ($\xi = 1.0$), similar to the elastoplastic yielding limit.}\label{Fig S4}     
\end{figure}

\begin{figure}[h]
    \centering
    \includegraphics[width=0.55\linewidth]{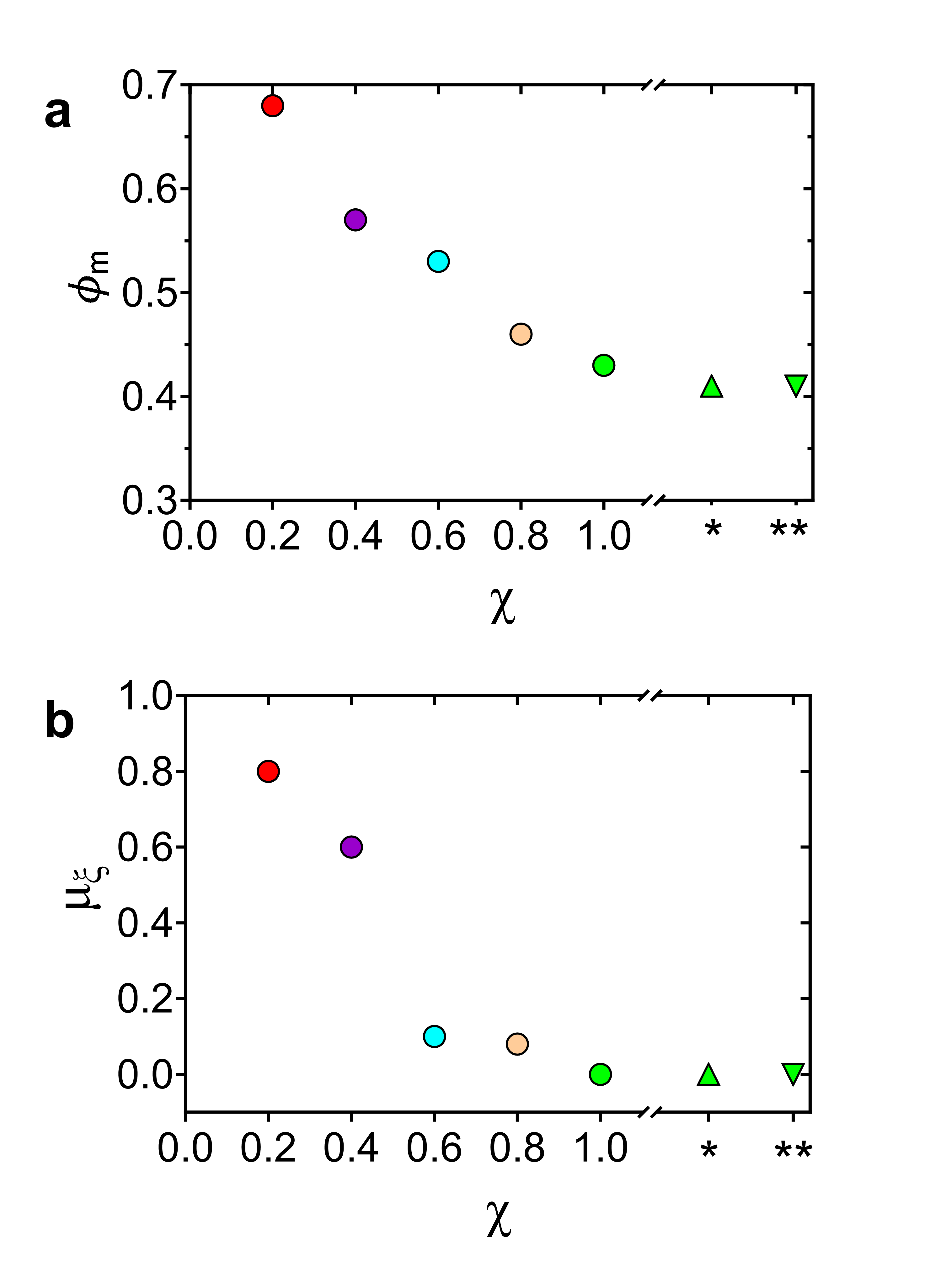}
    \caption{The fit parameters used for theoretical estimation of rearrangement time scale:(\textbf{a}) the suspension jamming point ($\phi_m$) and (\textbf{b)}) the average solid clay fraction-dependent friction coefficient ($\mu_\xi$), plotted as functions of increasing attraction parameter, $\chi = \chi(\xi,\zeta)$, in the Soft Earth suspensions. Fit values show a decrease in both the parameters as the clay fraction increases. The fitted friction parameter drops rapidly with increasing clay concentration in the mixtures, and disappears at $\xi = 1.0$.}\label{Fig S5}     
\end{figure}

\end{document}